\documentclass[
reprint,
amsmath,amssymb,
prb,
eprint,
floatfix,
]{revtex4-2}

\usepackage{graphicx,xcolor}% Include figure files
\usepackage{dcolumn}% Align table columns on decimal point
\usepackage{bm}% bold math
\usepackage[colorlinks,linkcolor=blue,citecolor=blue,urlcolor=blue]{hyperref}

\begin{document}

\preprint{APS/123-QED}

\title{Tuning the Intrinsic Spin Hall Effect by Charge Density Wave Order\protect\\
in Topological Kagome Metals
}

\author{Diana Golovanova}
\author{Hengxin Tan}%
 \author{Tobias Holder}%
\author{Binghai Yan}%
\email{binghai.yan@weizmann.ac.il}
\affiliation{%
    Department of Condensed Matter, Weizmann Institute of Science, Rehovot 7610001, Israel
}%
\date{\today}% It is always \today, 
\begin{abstract}
Kagome metals are topological materials with a rich phase diagram featuring various charge density wave orders and even unconventional superconductivity.
However, little is still known about possible spin-polarized responses in these non-magnetic compounds.
Here, we perform ab-initio calculations of the intrinsic spin Hall effect (SHE) in the kagome metals AV$_3$Sb$_5$ (A=Cs, Rb, K), CsTi$_3$Bi$_5$ and ScV$_6$Sn$_6$. We report large spin Hall conductivities, comparable with the Weyl semimetal TaAs. Additionally, in CsV$_3$Sb$_5$ the SHE is strongly renormalized by the CDW order. 
We can understand these results based on the topological properties of band structures, demonstrating that the SHE is dominated by the position and shape of the Dirac nodal lines in the kagome sublattice.
Our results suggest kagome materials as a promising, tunable platform for future spintronics applications.
\end{abstract}

%\keywords{Suggested keywords}%Use showkeys class option if keyword
%display desired
\maketitle

\section{\label{sec:level1}Introduction}

In recent years, a new class of topological materials, known as kagome metals, has drawn significant attention due to their unique band structure which combines Dirac point, van Hove singularity, and flat bands~\cite{jiang2023kagome,neupert2022charge}.
Owing to the geometric frustration~\cite{Zhao2021cascade, cho2021emergence, liu2021charge, luo2022electronic}, these compounds host a variety of exotic phenomena, such as the quantum spin liquid phase~\cite{yan2011spin,han2012fractionalized}, superconductivity~\cite{ko2009doped}, Weyl semimetal~\cite{Yang_2017,liu2019magnetic,morali2019fermi}, and anomalous Hall effect~\cite{nakatsuji2015large,nayak2016large,liu2018giant}.
Very recently, a charge density wave (CDW) order has been discovered in the non-magnetic $Z_2$ kagome metals AV$_3$Sb$_5$ (A = K, Rb, Cs)~\cite{ortiz2019new} at 78$\sim$102 K which is intertwined with a  superconducting state at about 0.9$\sim$2.5 K \cite{ortiz2020cs,ortiz2021superconductivity,Yin_2021}.
The CDW is closely linked to the electronic structure and Fermi surface of AV$_3$Sb$_5$ \cite{ortiz2020cs,tan2021charge,ortiz2021fermi,Zhao2021cascade}, and has been connected with a variety of other phenomena, such as electronic nematicity \cite{nie2022charge,jiang2023observation,xu2022three,li2022rotation}, quantum oscillations \cite{ortiz2021fermi,fu2021quantum,broyles2022effect,tan2023emergent}, and possible time-reversal symmetry breaking \cite{jiang2021unconventional,yang2020giant,yu2021concurrence,Kenney2021absence,FENG20211384,li2022no,mielke2022time}.
A pair density wave order in the superconducting phase is also observed \cite{chen2021roton}.
Besides, upon hole doping with Ti, recent experiments found that the CDW is suppressed in the sister compound CsTi$_3$Bi$_5$ \cite{yang2022titanium2}, which provides a platform to compare the consequence of CDW in these materials.
In addition to the extensively studied AV$_3$Sb$_5$-related materials, the CDW has been also discovered very recently in a bilayer kagome metal ScV$_6$Sn$_6$ at 92 K~\cite{Arachchige2022charge}. Distinctly different from the CDW in AV$_3$Sb$_5$ which mainly involves a distortion of the kagome sublattice, the CDW in ScV$_6$Sn$_6$ is originating rather from the non-kagome sublattices, with only very limited electronic structure changes~\cite{tan2023PRL}. 

This varied phenomenology of the CDW ordering in kagome metals calls for a detailed investigation of the commonalities and differences in the electronic structure of these compounds.
Here, we approach this question by investigating the nature, strength and chemical dependence of the spin Hall effect for several kagome materials. 
The spin Hall effect (SHE)~\cite{Sinova2015} describes the creation of a transverse spin current when a charge current is driven in longitudinal direction, which has been studied intensely in recent years for its potential applications in spin torque devices and related spintronics applications~\cite{Manchon2019}.
Nonetheless, so far not many nonmagnetic materials have been reported which support large spin polarized current responses.
The intrinsic SHE is a sensitive probe of the electronic structure which is enhanced in a number of topological insulators~\cite{Pesin2012,Zhou2015,Fan2016}, Weyl semimetals~\cite{sun2016strong,Smejkal2018} and nodal line semimetals~\cite{ sun2017dirac, hou2021prediction, yi2023topological}.
Specifically, it has been suggested that the SHE benefits from the presence of Dirac nodal lines~\cite{sun2017dirac}, making kagome materials a strong candidate platform for spin-polarized responses.

In the following, we will focus on the characterization and prediction of the SHE in CsV$_3$Sb$_5$ and related compounds using 
\emph{ab-initio} calculations. 
Most notably, we report a sizable SHE of almost $600 (\hbar/e) (\Omega \text{cm})^{-1}$ for pristine phase considering the in-plane component of the conductivity tensor $\sigma_{xy}^z$.
As the origin of intrinsic SHE we identify the Dirac nodal loops which are gapped out by spin-orbit coupling (SOC). 
The Berry curvature is spread smoothly in the vicinity of these gapped nodal lines, thereby leading to the large SHC.
We find that the spin conductivity is strongly suppressed in the CDW phase and additionally also sensitive to the type of CDW ordering. 
For the kagome metal ScV$_6$Sn$_6$,we report a large out-of-plane component of the conductivity tensor with an in-plane spin polarization. Specifically, ScV$_6$Sn$_6$ achieves a value of almost $560 (\hbar/e) (\Omega \text{cm})^{-1}$ for $\sigma_{zx}^y$ and $\sigma_{yz}^x$ near the Fermi energy. Surprisingly, in the latter compound the SHE in is only very weakly affected by the CDW transition.

We also briefly discuss CsTi$_3$Bi$_5$, which is closely related to CsV$_3$Sb$_5$ by chemical substitution. Since the Fermi level is located further way from the Dirac crossings and Van-Hove singularities in this compound, the SHC turns out to be much smaller.

The paper is structured as follows. In Sec. II, we introduce the main ingredients of the numerical approach. We present our findings for CsV$_3$Sb$_5$ in the pristine and the CDW ordered phases in Sec. III, followed by a comparison to other kagome compounds which further highlights the sensitivity of the SHE in kagome metals. We conclude in Sec. IV.

\begin{figure*}
\includegraphics[width=0.95\linewidth]{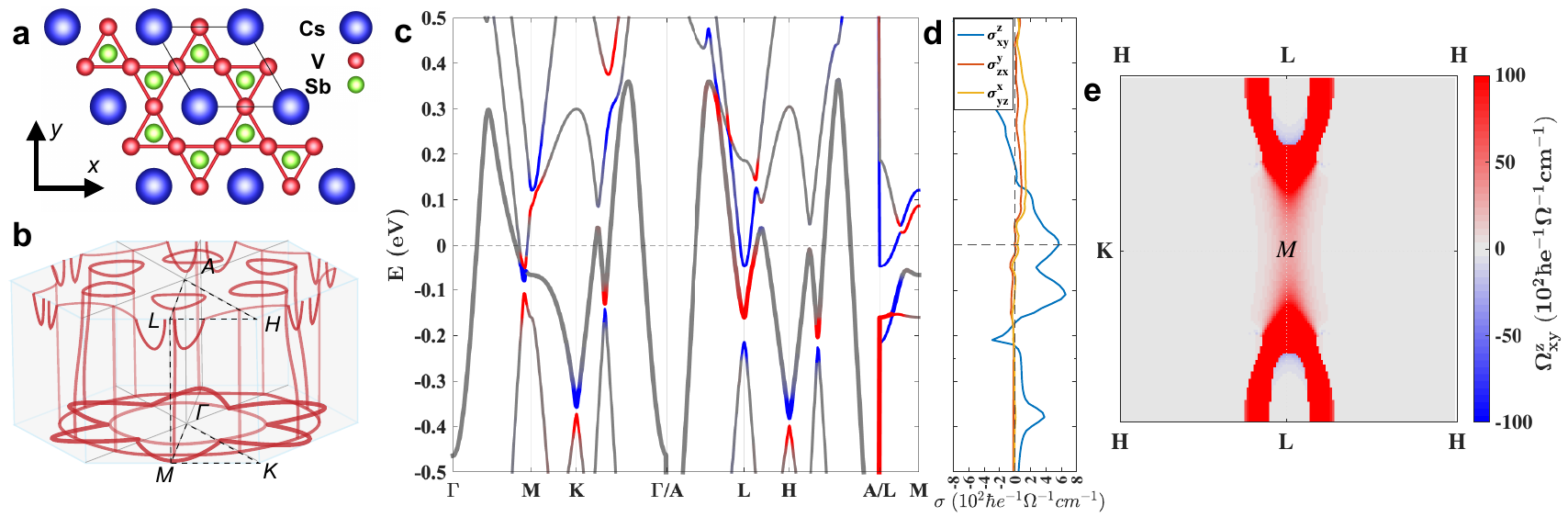}
\caption{\label{fig:one} Spin Hall Conductivity in CsV$_3$Sb$_5$.(a) Crystal structure of CsV$_3$Sb$_5$ in the $xy$-plane, depicting four unit cells. 
(b) Position of the Dirac nodal lines which are gapped due to SOC. Depicted is the nodal structure for the band crossing the Fermi level (band highlighted in (c)). All Dirac nodal lines reside in mirror planes.
(c) Band structure of CsV$_3$Sb$_5$, the coloring indicates regions with a large spin Berry curvature. Note that all bands are doubly degenerate, with each one carrying opposite Berry curvature, making the material time-reversal symmetric. The k-path of the band structure is indicated by a dotted line in (b). The color bar is the same as in (e).
(d) All three non-zero components of the spin Hall conductivity along the major crystal axes. $\sigma^z_{xy}$ has by far the largest value. (e) Map of the spin Berry curvature along the MKHL boundary plane of the Brillouin zone integrated until the Fermi level. This cut makes the largest contribution to the SHC peak at zero energy.
}
\label{fig:CVS_cond}
\end{figure*}

% \section{Model}
% \textit{The compound CsV$_3$Sb$_5$.}

\section{Method}
\textit{Intrinsic Spin-Hall Conductivity.} The spin Hall effect has two mechanisms for inducing spin current. The first one is the extrinsic spin Hall effect, which appears due to the interaction of carriers with local magnetic moments. For oppositely directed spins, this leads to the appearance of a Lorentz force in opposing directions, thus a spin Hall current. 
The second mechanism is intrinsic, which means it stems from the geometric phases imposed by the spin Berry curvature $\Omega^{\alpha}$ of the band structure, where $\alpha$ denotes the spin quantization axis.
For a tight-binding Hamiltonian with eigenvalues $E_{n,\bm{k}}$ and Bloch wavefunctions $|u_{n,\bm{k}}\rangle$, where $n$ denotes the band index, the band resolved spin Berry curvature is given by
\begin{equation} \label{eq:berry}
\begin{aligned}
\Omega_{n, i j}^{\alpha}(\bm{k}) & =-2 \operatorname{Im} \sum_{m \neq n} \frac{\langle u_{n, \bm{k}} |\hat{J}_i^{\alpha}| u_{m, \bm{k}}\rangle\langle u_{m, \bm{k}}|\hat{v}_j| u_{n, \bm{k}}\rangle}{\left(E_{n, \bm{k}}-E_{m, \bm{k}}\right)^2}.
\end{aligned}
\end{equation}
Here, we introduced the spin current operator $\hat{J}_i^{\alpha}=\frac{1}{2}\{\hat{v}_i, \hat{S}_{\alpha}\}$, the spin projection operator $\hat{S}_{\alpha}$ and the velocity operator $\hat{v}_i = \hbar^{-1} ( \partial H / \partial k_i)$. The indexes $i, j, \alpha$ denote projections along the spatial directions $x, y, z$. Summing over $\Omega^{\alpha}$ for all occupied states yields the spin Hall conductivity $\sigma_{i j}^{\alpha}$ in linear response, i.~e.
\begin{equation} \label{eq:conduct}
\begin{aligned}
\sigma_{i j}^{\alpha} & =e \hbar \int_{B Z} \frac{d^3 k}{(2 \pi)^3} \sum_n f_{n} (\bm{k}) \Omega_{n, i j}^{\alpha}(\bm{k}),
\end{aligned}
\end{equation}
where $f_{n} (\vec{k})$ is the occupation function. 
The method of estimation of SHE in the materials using the formula \ref{eq:conduct} is widely employed.\cite{guo2008intrinsic, sun2016strong, sun2017dirac, hou2021prediction, yi2023topological}

\textit{DFT calculations.}
The spin Berry curvature and spin Hall conductivity are evaluated by employing the tight-binding Hamiltonians, which are extracted from the density functional theory (DFT) calculations for all materials.
All DFT calculations are performed with the full-potential local-orbital minimum-basis code (FPLO) \cite{FPLO}.
The default atomic basis sets are employed for the wave function expansion. The Perdew-Burke-Ernzerhof-type generalized gradient approximation \cite{PEB} is used to describe the exchange-correlation interaction.
$k$-meshes of 12$\times$12$\times$6 and 6$\times$6$\times$6 are used to sample the Brillouin zones (BZ) of the pristine and CDW phases of CsV$_3$Sb$_5$, respectively. The crystal structures of CsV$_3$Sb$_5$ and ScV$_6$Sn$_6$ are obtained respectively from Ref.~\onlinecite{tan2021charge,tan2023PRL}.
The tight-binding Hamiltonians of all phases are extracted via the maximally localized Wannier functions \cite{wannier} as implemented in FPLO, which enforces all crystal symmetries.
The Wannier basis set for CsV$_3$Sb$_5$ is composed of the V $d$ and Sb $p$ orbitals, and composed of V $d$ and Sn $p$ orbitals for SnV$_6$Sn$_6$.
To calculate the SHC, the momentum integration was done with a $k$-mesh of $200\times200\times200$, which reliably captures the Fermi surface topology. 

\section{\label{sec:level2}Results}
\subsection{\label{sec:sublevel2.1}Spin Hall Conductivity in CsV$_3$Sb$_5$}

AV$_3$Sb$_5$ settles into a high-symmetry phase, crystallizing within the $P6/mmm$ (No. 191) space group. The structure consists of a kagome sublayer of Vanadium (V), with Antimony (Sb) positioned at the center of the hexagon. Additionally, there's a honeycomb sublayer of Sb and a triangular sublayer composed of either Cesium (Cs), Rubidium (Rb), or Potassium (K)~\cite{ortiz2019new}. The structure of CsV$_3$Sb$_5$ viewed from the top is depicted in Fig.~\ref{fig:CVS_cond}a. 
Due to structural properties CsV$_3$Sb$_5$ preserves numerous symmetries. Mirror planes provide this material with a dense structure of Dirac nodal lines, as shown in Fig.~\ref{fig:CVS_cond}b. 
The resulting band structure of CsV$_3$Sb$_5$ with SOC in Fig.~\ref{fig:CVS_cond}c has a parabolic band near the $\Gamma$-point originating from the Sb $p_z$ orbital. The remaining bands near the Fermi level mostly carry the orbital character of V $d$ orbitals~\cite{wang2022origin}, which highlights the importance of the kagome lattice for CsV$_3$Sb$_5$. Furthermore, AV$_3$Sb$_5$ preserves time reversal and inversion symmetries, leading to double band degeneracy. They correspond to spin-up and down polarized states.

A calculation of the band- and momentum-resolved spin Berry curvature using Eq.~\ref{eq:berry} reveals the location of the Berry charges, as indicated by the red and blue coloring in Fig.~\ref{fig:CVS_cond}. 
Clearly, the Berry charges are highly localized near (avoided) Dirac crossings, with opposite signs between the lower and upper branches in each crossing. Due to this antipodal appearance of the Berry charges, only Dirac crossings close to the Fermi level can contribute to the response.
We emphasize that the band structure is doubly degenerate due to PT symmetry. It results in the spin Berry curvature, the same for both degenerate PT partners.

In terms of linear response theory, the space group of kagome material CsV$_3$Sb$_5$ leads to another important consequence: The spin polarization, spin current and charge current should be mutually orthogonal. In this way, most entries of the spin conductivity tensor are identical zero and only six components are finite: $\sigma_{x y}^z=-\sigma_{y x}^z, \sigma_{z x}^y=-\sigma_{z y}^x$ and $\sigma_{y z}^x=-\sigma_{x z}^y$~\cite{kleiner1966space, seemann2015symmetry}.

For the numerical evaluation of the SHC, we used a regularized expression for the singular denominator in Eq.~\eqref{eq:berry}, replacing $(E_{n,\bm{k}}-E_{m,\bm{k}})^2\to (E_{n,\bm{k}}-E_{m,\bm{k}})^2+\eta^2$, where $\eta=10^{-3}$.
Fig.~\ref{fig:CVS_cond}d shows the resulting three non-vanishing components of the spin conductivity tensor $\sigma_{x y}^z$, $\sigma_{z x}^y$ and $\sigma_{y z}^x$. 
At the chemical potential, by far the largest component is $\sigma_{x y}^z$, with a value of $572 (\hbar/e) (\Omega \text{cm})^{-1}$. 
As a function of filling,  even slightly larger values can be achieved,  up to $658 (\hbar/e) (\Omega \text{cm})^{-1}$ at $\mu=-0.11$ eV. 
This is comparable to the values reported for Weyl semimetals, for example TaAs is believed to have a SHC of around $780 (\hbar/e) (\Omega \text{cm})^{-1}$~\cite{sun2016strong}.
The values for the other components are negligibly small.
Note that the spin conductivity has the unit $e/2\pi$, but we quote numerical values in terms of the charge conductivity times $\hbar/e$ for ease of comparison with the regular conductivity $\sigma$.

As the main source of non-zero Spin Berry curvature we identify the Dirac nodal lines which are weakly gapped due to SOC. This mechanism is similar to the ones demonstrated previously for a number of other materials~\cite{sun2016strong, sun2017dirac, hou2021prediction, yi2023topological}. In the case of CsV$_3$Sb$_5$, the nodal line structure stretches across the entire Brillouin zone, forming a dense cage of parallel lines both in-plane and along the z-axis. 
We point out that all nodal lines lie within the mirror planes of the Brillouin zone~\cite{sun2017dirac}, of which there are 8 in total.
The (avoided) Dirac crossings are most easily detected by considering the band structure without SOC, where the bands do cross.
The relevant nontrivial band crossings with singular Berry curvature in the 3d Brillouin zone are shown in Fig.~\ref{fig:CVS_cond}b. 
It is straightforward to see that the nodal lines predominantly constitute the SHC by comparing their positions in momentum space with the momentum-resolved spin Berry curvature of the respective band. 
As an example for this, we show in Fig.~\ref{fig:CVS_cond}e a two-dimensional cut of the spin Berry curvature along the vertical BZ boundary plane which constitutes the largest integration weight to the SHE at the Fermi level. 

\begin{figure*}
\includegraphics[width=0.85\linewidth]{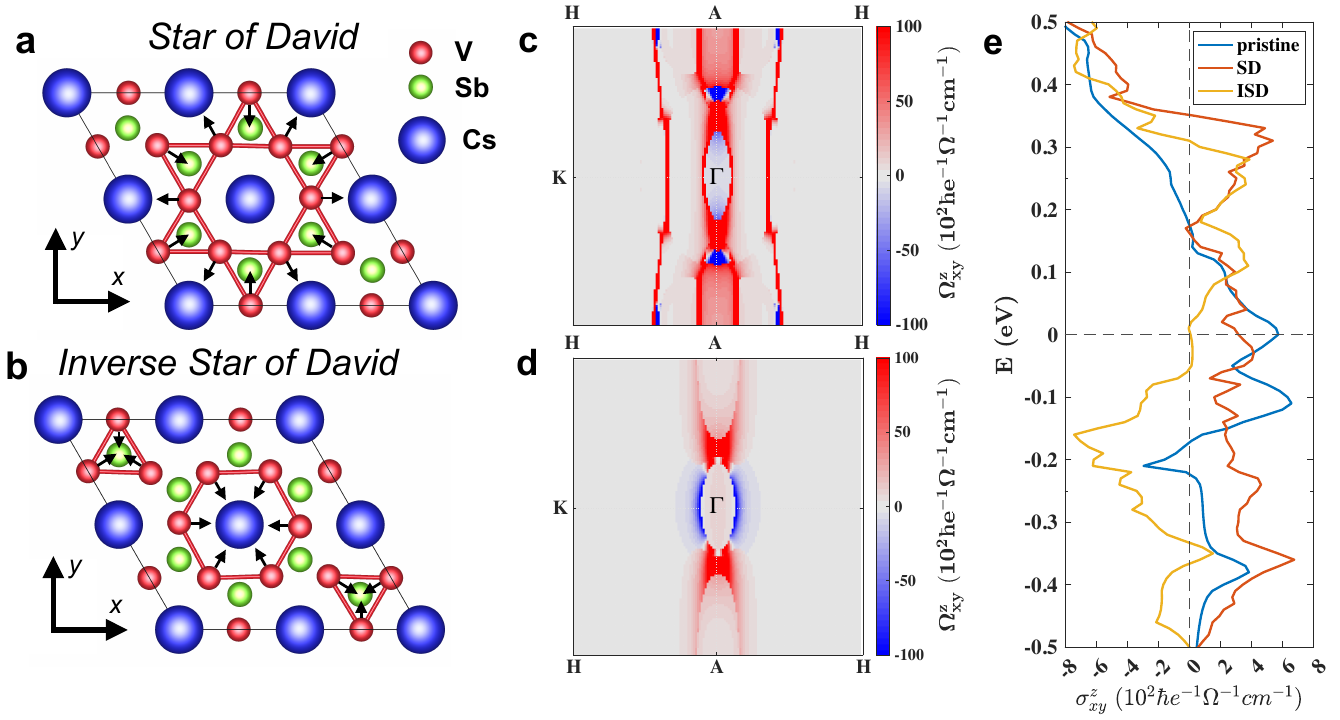}% Here is how to import EPS art
\caption{ \label{fig:CDW} Spin Hall Conductivity with Charge Density wave. (a,b) Both in-plane $2\times2$ charge density waves of the kagome metal CsV$_3$Sb$_5$: Star of David (SD) and Inverse Star of David (ISD). The black arrows show the directions of deformation for Vanadium atoms with respect to the pristine phase. (c,d) Map of the spin Berry curvature along the $\Gamma$KHA cut of the Brillouin zone integrated until Fermi level for SD (c) and ISD (d) phases. Note that the $MKHL$-plane from Fig~\ref{fig:CVS_cond}(e) folds onto the $\Gamma A$ line, with the corresponding integration weight being spread out to closeby momenta.  
Thus the SHC is mostly constituted by the contributions close to $\Gamma A$ in the folded BZ. 
(e) Comparison of the maximum spin hall components for all three phases: pristine, SD and ISD.}
\end{figure*}

Attempting to match the crossing points in Fig.~\ref{fig:CVS_cond}c with the peaks in the SHC as a function of chemical potential in Fig.~\ref{fig:CVS_cond}d,  one can see that not all Dirac nodal lines contribute equally to the integrated SHC. 
Clearly, the precise weight of each crossing depends on details of the dispersion and also on the orbital character of the respective band crossing in relation to the component of the SHC. 
However, we want to emphasize specifically that the weak but nonzero dispersion of the nodal lines along the line direction averages of the Berry curvature located in the vicinity of the avoided crossings, and therefore attenuates the peaks in the SHC.
Only if the gap due to SOC is large enough compared to this dispersion, the chemical potential can reside within the gap along the entire line and the attenuation effect does not play a role. This mechanism competes with the fact that a smaller gap leads to larger values in the Eq.~\ref{eq:conduct} for calculating the SHC.
This latter case is true for the peak of the SHC at $E=0\,\mathrm{eV}$, which is almost entirely due to the Dirac crossings in the vicinity of the M and L points, which happen to also be the locations of the kagome-related Van Hove singularities.

\subsection{\label{sec:sublevel2.2}Spin Hall Conductivity in the Charge Density wave state}
We now turn towards the symmetry broken phase with commensurate CDW order.
In kagome metals of type AV$_3$Sb$_5$, the CDW does not violate the original rotational symmetry, and it can assume two different deformation patterns in the xy-plane, which lead to a 2x2 unit cell and are called Star David (SD) and Inverse Star David (ISD) (cf. Fig.~\ref{fig:CDW}a,b). 
The CDW transition has been experimentally observed and occurs at temperatures between 78-102 K~\cite{ortiz2020cs, ortiz2021superconductivity, yin2021superconductivity}.

In the CDW state, the Brillouin zone has a size 1/2 x 1/2 in $xy$ plane.
This means that the bands go through a band-folding procedure and high-symmetry points like $M$ and $L$ will be transferred to the $\Gamma$ and $A$ points, respectively. The same applies for $K$ and $H$. 
In the folded Brillouin zone it is much harder to locate and identify nodal lines. For this reason, we instead track the changes to spin Hall Berry curvature along the $MKHL$-plane cut which contributes most to the SHE in the pristine phase. 
After folding, this plane collapses onto the new $\Gamma A$-line, with most of the integration weight being redistributed to closeby momenta. Indeed, as clearly visible in the plane cuts along $\Gamma KHA$ in Fig.~\ref{fig:CDW}c,d, the largest contributions reside in close vicinity to  the $\Gamma A$-line. The band structure along the characteristic 1D path for the two CDW phases is presented in the appendix~\ref{app:A}.

Fig.~\ref{fig:CDW}e shows 
the intrinsic spin Hall conductivity (SHC) according to Eq.~\ref{eq:conduct} for SD and ISD charge density waves. We note in passing that the result for the $2\times 2 \times 2$ CDW, which is composed of SD and ISD stacked one above the other, essentially falls in between either planar CDW. Details of the latter can be found in appendix~\ref{app:A}. 
The SHC as a function of chemical potential is qualitatively different between both CDW orders and also between the CDWs and the pristine phase.
The SD phase yields a spin Hall conductivity of $\sigma_{xy}^z = 316 (\hbar/e) (\Omega \text{cm})^{-1}$, while in contrast, the ISD suppresses it almost completely.
In short, both CDW types decrease the spin Hall effect. 
Based on the momentum cuts, we conclude that the much smaller value for the ISD state is due to an overall reduction of the spin Berry curvature and additionally due to partial cancellations.
It is known that the CDW in CsV$_3$Sb$_5$ opens partial gaps close to the $M$ and $L$ points, which substantially decreases the density of states~\cite{tan2021charge,luo2022electronic}. We attribute the suppression of the SHE to these dramatic changes of the band structure. 

The suppression of the SHE can be recreated in a simplified three-band kagome model.  To this end, note that it 
is possible to identify in the band structure of CsV$_3$Sb$_5$ two characteristic band patterns of the kagome lattice~\cite{wu2021nature}. 
One of these kagome patterns has a Van Hove singularity close to the Fermi level. 
As we show in appendix~\ref{app:B}, for such a filling, both the SD and the ISD order substantially reduce the SHC, rendering it completely zero for the ISD phase, which becomes fully gapped in the three-band model. 

In conclusion, both the suppression of the spin Berry curvature and the suppression of the density of states contribute to the strong modification of the SHE between the pristine phase and the ordered state.

\subsection{\label{sec:sublevel2.3}Comparison to other Kagome materials}

\paragraph{AV$_3$Sb$_5$.} 

\begin{figure}
\includegraphics[width=0.7\linewidth]{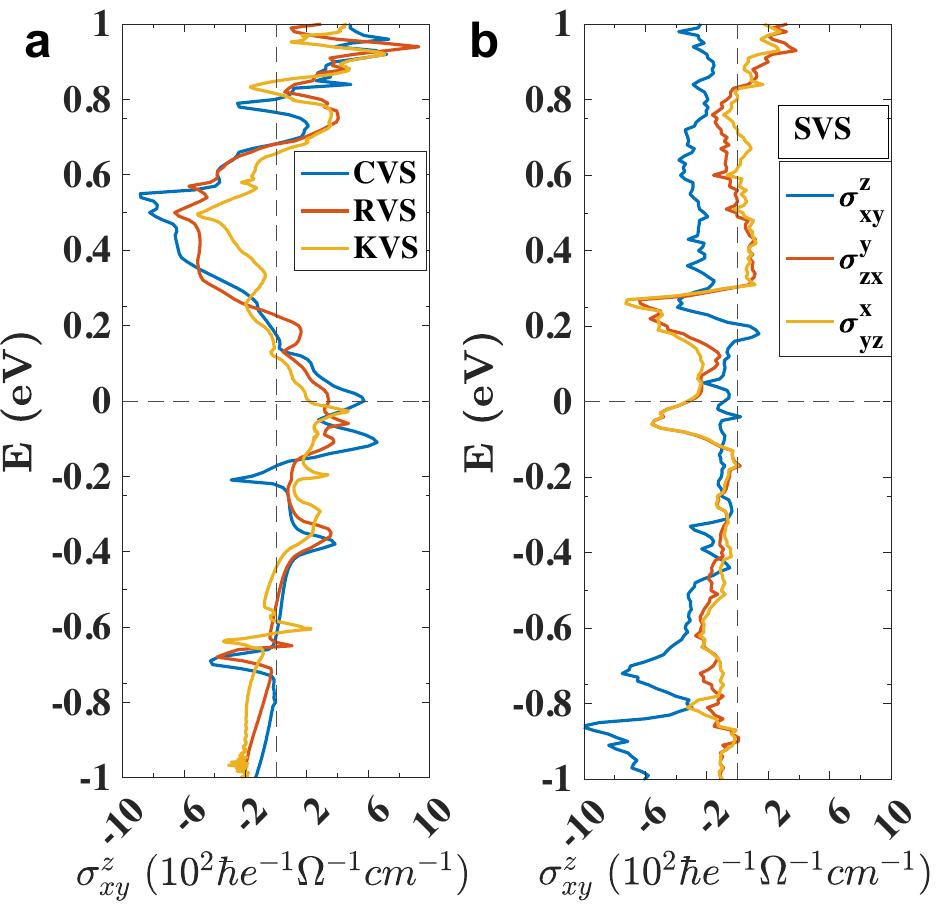}% Here is how to import EPS art
\caption{\label{fig:comp_comp} Comparison of SH conductivities of different kagome metals. (a) SHC for materials from the AV$_3$Sb$_5$ family (A=Cs, Rb, K). (b) Components of the SHC tensor of ScV$_6$Sn$_5$. }
\end{figure}

Many of the arguments applicable to CsV$_3$Sb$_5$ are also applicable to KV$_3$Sb$_5$ and RbV$_3$Sb$_5$: they have a common phenomenology, similar band structures and the same number of valence electrons. The only difference is which heavy atom is placed into the triangular sublattice, which affects the mechanism of spin-orbit interaction between different orbitals. In Fig.~\ref{fig:comp_comp}, we see a trend that an atom with a smaller mass is characterized by a smaller conductivity amplitude, which agrees with the classical intuition that lighter atoms cause less spin-orbit interaction. However, while CsV$_3$Sb$_5$ exhibits the largest value of $\sigma_{xy}^z$ at the Fermi energy; this is not the case for other energy levels. In fact, any of the three sister compounds can exhibit the largest SHC, given a suitable chemical potential. This indicates that the value of the chemical potential is more crucial in determining the best candidate for demonstrating the SHE. Other components of the conductivity tensor in the K- and Rb- compounds are small similar to the case of CsV$_3$Sb$_5$.

In other AV$_3$Sb$_5$ materials we expect that the formation of the same CDW order leads to a similar suppression of the SHC compared to the pristine state. 
The reason for this is twofold. A reduction of the density of states similar to the one in CsV$_3$Sb$_5$ has been reported both in in RbV$_3$Sb$_5$ \cite{cho2021emergence} and KV$_3$Sb$_5$ \cite{jiang2021unconventional}, Secondly,  the simplified kagome model discussed in Appendix~\ref{app:B} shows that the suppression is generic for this particular CDW order. 
Thus, it is to be expected that CDW formation causes the suppression of SHC value near the Fermi level in all members of the family of AV$_3$Sb$_5$ materials.

\paragraph{CsTi$_3$Bi$_5$.}

\begin{figure}
\includegraphics[width=1.01\linewidth]{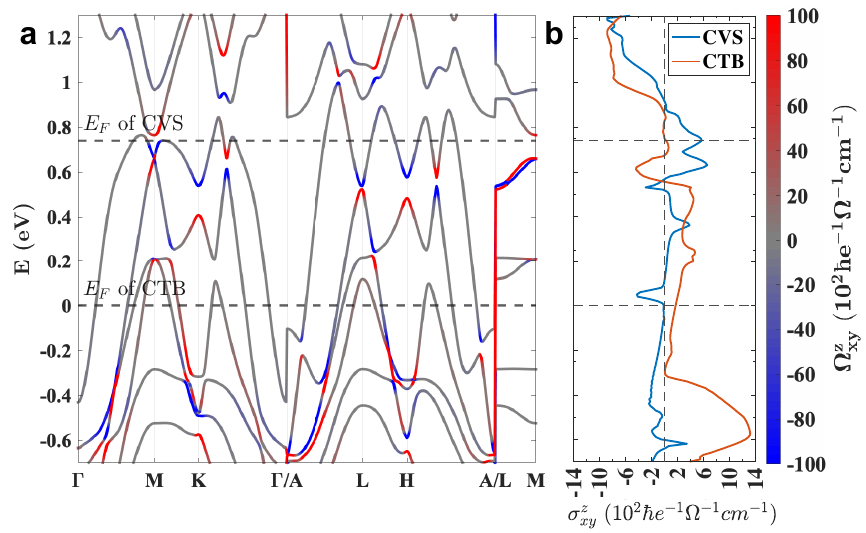}% Here is how to import EPS art
\caption{\label{fig:compCTB} Comparison of the spin conductivities of two kagome metals: CsV$_3$Sb$_5$ and CsTi$_3$Bi$_5$. (a) The band structure of CsTi$_3$Bi$_5$ with band-resolved spin Berry curvature. For comparing the conductivities, the CsV$_3$Sb$_5$ Fermi level is also shown. (b) Comparison of CsV$_3$Sb$_5$ and CsTi$_3$Bi$_5$ SHC.
}
\end{figure}

Since the large value of the SHC in CsV$_3$Sb$_5$ arises almost entirely from the bands related to the kagome lattice of Vanadium atoms, this poses the interesting question whether the effect can be further amplified when replacing Vanadium by a heavier atom like Titanium(Ti), which also carries a larger intrinsic SOC. 
Fortunately, CsTi$_3$Bi$_5$ has been recently synthesized in the same lattice configuration, which makes it a suitable candidate for comparison.

The main chemical difference between both materials is the number of valence electrons.
Namely, Ti has one less valence electron compared to V, which lowers the position of the chemical potential in CsTi$_3$Bi$_5$.
We point out that no CDW formation could be detected at all in CsTi$_3$Bi$_5$~\cite{li2022electronic}. 
The ab-initio band structure of CsTi$_3$Bi$_5$ is depicted in Fig.~\ref{fig:compCTB}\text{a}, with the corresponding SHC shown in Fig.~\ref{fig:compCTB}\text{b}. 
In order to compare both compounds, we reconstruct the respective Fermi energy of  CsV$_3$Sb$_5$ by adding three electrons to the integrated density of states in CsTi$_3$Bi$_5$, which leads to the reconstructed energy $E=0\,\mathrm{eV}$ in CsTi$_3$Bi$_5$. Based on this reference point, it is possible to also overlay the SHC curves of both materials.
While the band structures of both materials do not match closely, the energy dependence of the SHC exhibits the same overall shape.

However, contrary to expectations $\sigma^z_{xy}$ does not increase much.  Particularly near the reconstructed Fermi energy of CsV$_3$Sb$_5$, the corresponding value for CsTi$_3$Bi$_5$ is practically negligible.
The reason is that even if the positions of the Dirac nodal lines have changed only slightly, this small change dramatically affects the spin conductivity. 
More seriously, at the actual Fermi level of CsTi$_3$Bi$_5$, the SHE is still relatively small due to the comparatively small number of (avoided) Dirac crossings which can be found in the vicinity of the new Fermi level.

This comparison highlights potential shortcomings of overly simplistic models of the spin Hall effect. 
In particular for kagome metals,
we conclude that an increase of the intrinsic SOC by chemical substitution is not very effective at increasing the spin Hall conductivity.

\paragraph{ScV$_6$Sn$_6$.}

\begin{figure*}
\includegraphics[width=0.89\linewidth]{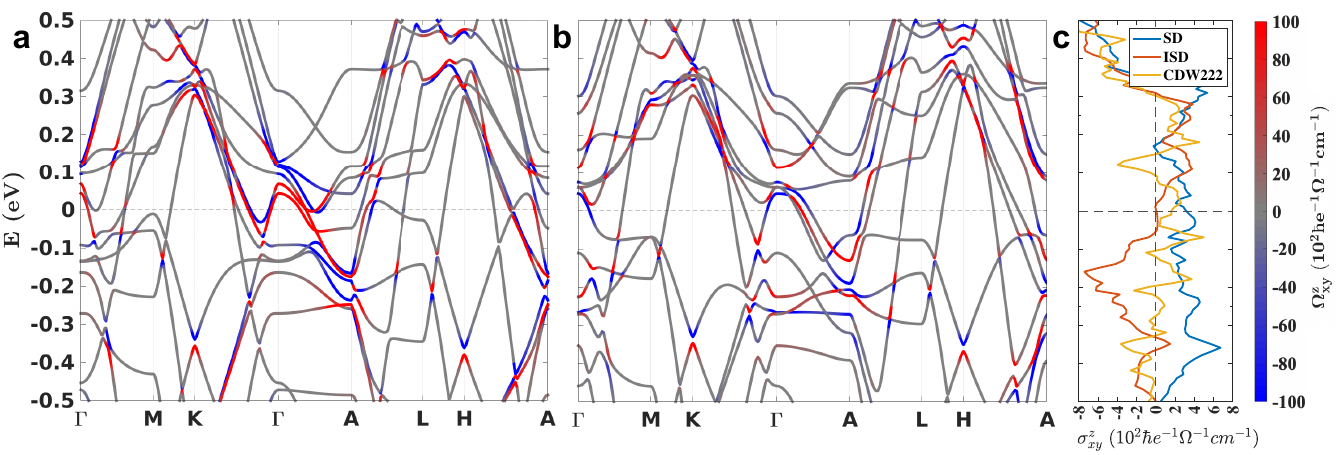}% Here is how to import EPS art
\caption{\label{fig:band_CVS_CDW} Comparison of two CDW band structures with band resolved spin conductivity for CsV$_3$Sb$_5$: (a) Star of David. (b) Inverse Star of David. (c) Comparison of in plane component of SHC for three CDW phases: SD, ISD and $2\times 2 \times 2$ CDW, which is composed of two distorted kagome lattices, namely SD and ISD, stacked one above the other. }
\end{figure*}

Another promising family of kagome materials enriched by Dirac nodal lines is RV$_6$Sn$_6$ (R = Y, Gd-Tm, and Lu). We present the calculation of the SHC for the ScV$_6$Sn$_6$ compound in Fig.~\ref{fig:comp_comp}b. In contrast to AV$_3$Sb$_5$ and CsTi$_3$Bi$_5$ it exhibits larger values of the SHC tensor in directions orthogonal to the sublattice planes. The reasons for this behavior are likely the structural differences between these compounds. Notably, ScV$_6$Sn$_6$ comprises two adjacent Kagome sublattices of Vanadium atoms instead of a single one. This unique configuration nearly halves the distance between the Vanadium atoms.
Since the values of $\sigma_{zx}^y$ and $\sigma_{yz}^x$ in ScV$_6$Sn$_6$ are comparable with the values of $\sigma_{xy}^z$ in CsV$_3$Sb$_5$, this opens up the potential for utilizing this class of materials in practical applications. In the majority of spintronics devices, it's necessary for the spin current to flow in a direction perpendicular to the substrate plane( corresponding to $z$ direction in our notation).
We note that in strong distinction to the prominent role of the CDW in CsV$_3$Sb$_5$, the charge ordered state in ScV$_6$Sn$_6$ barely affects the electronic band structure, and has likewise essentially no effect on the SHE (cf. appendix~\ref{app:C}).

\section{Conclusion}
In this work, we studied the spin-polarized response in the recently synthesized family of kagome materials. Based on an \textit{ab-initio} approach we identified CsV$_3$Sb$_5$ and ScV$_6$Sn$_6$ as the most promising compounds and highlighted the presence of gapped Dirac nodal lines as the main mechanism for generating a Spin Hall conductivity in these materials. Using the examples of CsV$_3$Sb$_5$ and ScV$_6$Sn$_6$, we considered the interplay between SHE and CDW formation. For the former, the charge density order strongly affected the Spin Hall conductivity, even completely suppressing it for ISD ordering. In contrast, almost no changes were found in ScV$_6$Sn$_6$. 
These differences reinforce the notion that nodal lines are an important mechanism for the SHE, and show that the spin Hall conductivity can serve as a sensitive indicator of topological band crossings. 
Additionally, the large intrinsic spin Hall effect that we have reported here showcases the potential of kagome metals as a platform for spin selective  devices: Due to their metallic nature they can support currents much better than most other candidate materials which are typically gapped~\cite{Manchon2019}.

\begin{acknowledgments}
B.Y.\ acknowledges the financial support by the European Research Council (ERC Consolidator Grant No. 815869, ``NonlinearTopo'') and Israel Science Foundation (ISF No. 2932/21). 
\end{acknowledgments}

\appendix

\section{Details of the CDW in CsV$_3$Sb$_5$} \label{app:A}

The Charge Density Wave (CDW) phenomenon leads to a fourfold increase in the size of the unit cell, resulting in a denser band structure and the displacement of Dirac nodal lines due to band folding. In the newly defined Brillouin Zone (BZ), the majority of topological charges are concentrated along the $\Gamma A$ path, which is evident from the colored band structures depicted in Fig.~\ref{fig:band_CVS_CDW}. As was discussed in the main text, the ISD causes an almost complete suppression of SHE. The reason for this is also reflected in the 1d cut of BZ. The density of bands near the Fermi energy in the ISD order (Fig.~\ref{fig:band_CVS_CDW}b) is much lower than for the SD order (Fig.~\ref{fig:band_CVS_CDW}a). Furthermore, the spin Berry curvatures of the two nontrivial bands crossing at the Fermi level between $\Gamma$ and $A$ have opposite signs, thereby negating each other.

CsV$_3$Sb$_5$ also hosts a three-dimensional $2 \times 2 \times 2$ CDW, where planar distortions of Vanadium of SD and ISD are stacked on top of each other in an alternating fashion, thus increasing the unit cell by a factor of two also in the $z$ direction. The SHC tensor in the $xy$ plane for such CDW is shown in Fig.~\ref{fig:band_CVS_CDW}c), superimposed on the SD and ISD curves for comparison. As $2 \times 2 \times 2$ CDW is constituted by the SD and ISD lattices, its yellow curve predominantly falls between the red and blue ones. At the Fermi energy level, it adopts a value that is half of the SD one.

\section{Kagome model with CDW} \label{app:B}

\begin{figure}
\includegraphics[width=1.\linewidth]{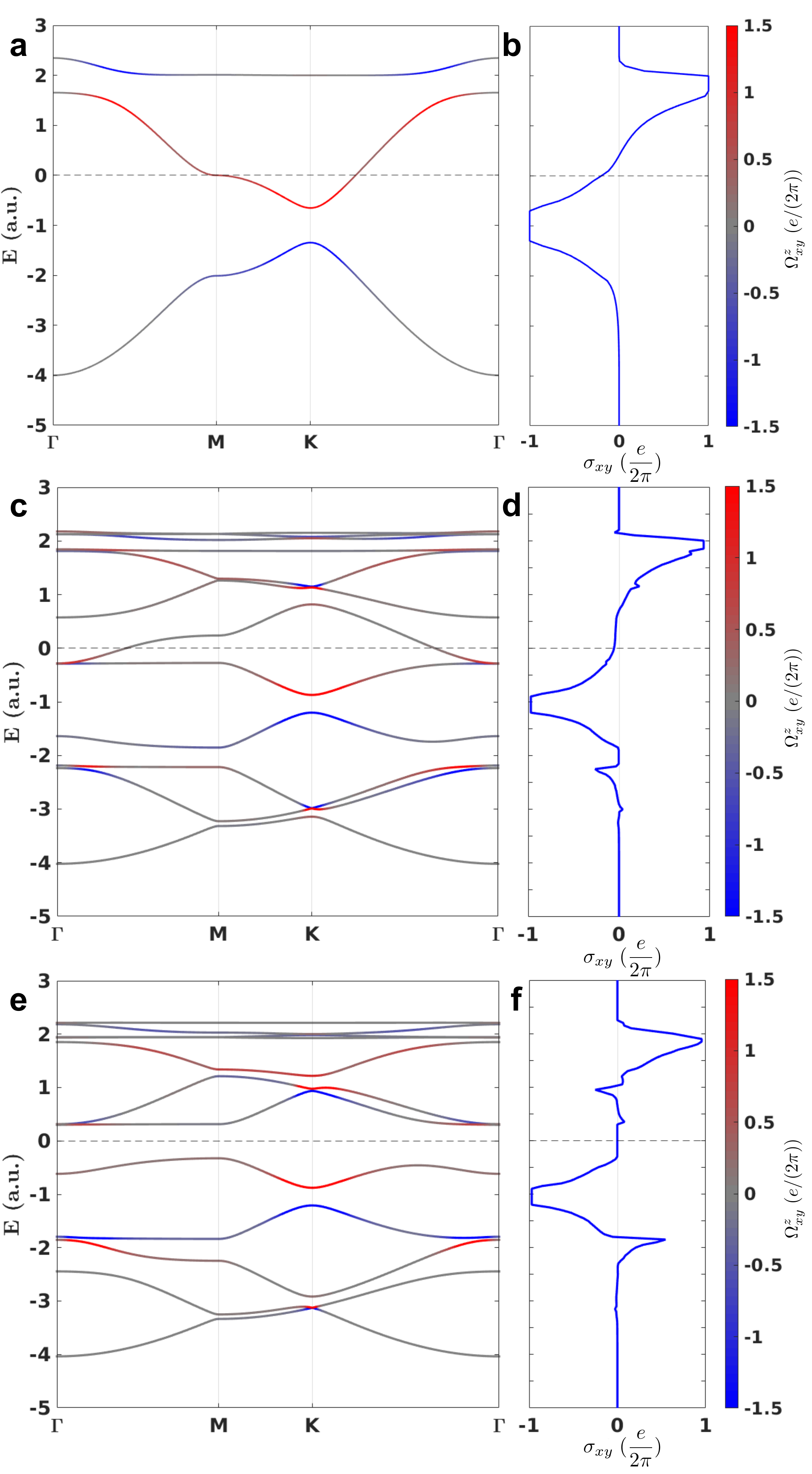}% Here is how to import EPS art
\caption{\label{fig:kagome} Spin Hall Conductivity in Kagome lattice. (a)-(b) Band structure and energy resolved SHC for pristine phase. The SOC parameter is $\lambda_{KM}=0.1$. (c)-(d) Band structure and energy resolved SHC for the Star of David (SD) phase. $\lambda_{KM}=0.05$. The value of CDW transformation for a hexagon is $110\%$ and $120\%$ for a triangle. (c)-(d) Band structure and energy resolved SHC for the Inverse Star of David (ISD) phase using $\lambda_{KM}=0.05$. The value of CDW distortion for SD (ISD) is $90\%$ ($80\%$). The coloring in the band structure indicates the momentum-resolved spin Berry curvature.}
\end{figure} 

Let's consider the tight-binding model for independent electrons on the 2D Kagome lattice. The spin-independent part of the Hamiltonian is given by 
\begin{equation}
H_0=t_0 \sum_{\langle i, j\rangle}\left(c_{i \alpha}^{\dagger} c_{j \alpha}+\text { h.c. }\right),
\end{equation}
where $t_{i j} = t_0$ is the hopping amplitude between nearest neighbor link $\langle i,j\rangle$ and $c_{i \alpha}^{\dagger}$ ($c_{i \alpha}$) is the creation (annihilation) operator, $\alpha$ corresponds to the spin degree of freedom. For simplicity, we take $t_0 = 1$ as the energy unit and distance $a=1$ between the nearest sites as the unit of length. 
This Hamiltonian reads in momentum space,
\begin{equation}
\mathcal{H}_0=\sum_{\mathbf{k}} \psi_{\mathbf{k}}^{+}\left[H_0(\mathbf{k}) \otimes \mathbf{I}_{2 \times 2}\right] \psi_{\mathbf{k}},
\end{equation}
where the $2 \times 2$ unit matrix $\mathbf{I}_{2 \times 2}$ denotes the spin degeneracy and $\psi_{\mathbf{k}}=\left(c_{A \mathbf{k} \uparrow}, c_{B \mathbf{k} \uparrow}, c_{C \mathbf{k} \uparrow}, c_{A \mathbf{k} \downarrow}, c_{B \mathbf{k} \downarrow}, c_{C \mathbf{k} \downarrow} \right)^{T}$ is the six-component electron field operator, which includes three lattice sites in Wigner-Seitz unit cell and spin degree of freedom. $H_0(\mathbf{k})$ is defined as
\begin{equation}
H_0(\mathbf{k})=\left(\begin{array}{ccc}
0 & 2 \cos \left(\mathbf{k} \cdot \mathbf{a}_1\right) & 2 \cos \left(\mathbf{k} \cdot \mathbf{a}_3\right) \\
2 \cos \left(\mathbf{k} \cdot \mathbf{a}_1\right) & 0 & 2 \cos \left(\mathbf{k} \cdot \mathbf{a}_2\right) \\
2 \cos \left(\mathbf{k} \cdot \mathbf{a}_3\right) & 2 \cos \left(\mathbf{k} \cdot \mathbf{a}_2\right) & 0
\end{array}\right)
\end{equation}
where $\mathbf{a}_1 = a(1, 0)^{T}$, $\mathbf{a}_2 = a(-\frac{1}{2}, \frac{\sqrt{3}}{2})^{T}$, $\mathbf{a}_3 = a(-\frac{1}{2}, -\frac{\sqrt{3}}{2})^{T}$ represent the displacements in a unit call from A to B site, from B to C and from C to A, respectively. 

We add to the model the effect of spin-orbit coupling interaction (SOC) of Kane-Mele type 
\begin{equation}
H_{\mathrm{SOC}}=i \lambda_{\mathrm{KM}} \sum_{\langle i j\rangle} v_{i j} c_i^{\dagger} \sigma_z c_j \text {. }
\end{equation}
This intrinsic SOC preserves one spin projection ($\sigma_z$) and opens a gap at the Dirac points, which means that all bands remain doubly degenerate. In momnetum space, the KM Hamiltonian becomes,
\begin{equation}
H_{\mathrm{SOC}}=\sum_{\boldsymbol{k}} \psi_{\boldsymbol{k}}^{\dagger}\left(\begin{array}{cc}
2 \lambda_{\mathrm{KM}} \Gamma_{\mathrm{KM}}(\boldsymbol{k}) & 0 \\
0 & -2 \lambda_{\mathrm{KM}} \Gamma_{\mathrm{KM}}(\boldsymbol{k})
\end{array}\right) \psi_{\boldsymbol{k}},
\end{equation}

\begin{equation}
\begin{aligned}
& \Gamma_{\mathrm{KM}}(\boldsymbol{k})= \\
& \left(\begin{array}{ccc}
0 & i \cos \left(\boldsymbol{k} \cdot \boldsymbol{b}_1\right) & -i \cos \left(\boldsymbol{k} \cdot \boldsymbol{b}_3\right) \\
-i \cos \left(\boldsymbol{k} \cdot \boldsymbol{b}_1\right) & 0 & i \cos \left(\boldsymbol{k} \cdot \boldsymbol{b}_2\right) \\
i \cos \left(\boldsymbol{k} \cdot \boldsymbol{b}_3\right) & -i \cos \left(\boldsymbol{k} \cdot \boldsymbol{b}_2\right) & 0
\end{array}\right)
\end{aligned}
\end{equation}
where we used the notation of next-nearest vectors $\bm{b}_1=\bm{a}_2-\bm{a}_3$, 
$\bm{b}_2=\bm{a}_3-\bm{a}_1$, 
$\bm{b}_3=\bm{a}_1-\bm{a}_2$.

The energy-resolved dependence of SHC against the band structure for this simple model is presented in Fig.~\ref{fig:kagome}a,b. Note that the orientation of the kagome band structure is chosen such that it qualitatively resembles the kagome-related bands in AV$_3$Sb$_5$, with the flat band on the top
and a Fermi level $E=0$ which approximates the Fermi level in CsV$_3$Sb$_5$. 
In order to capture the CDW ordering, we need to increase the unit cell of the kagome model and recreate the  displacements of Vanadium atoms in the kagome sublattice, in a fashion similar to how it is depicted in Fig.~\ref{fig:CDW}a,b of the main text. 
Increasing the size of the unit cell by $2 \times 2$ and then disproportionating the lattice distances accordingly, one can mimick the formation of SD and ISD distortions. The effects of this modulation on the band structure are shown in Fig.~\ref{fig:kagome}c-f. The most interesting result for the ISD state Fig.~\ref{fig:kagome}e,f is the complete opening of the gap close to the Van Hove singularity. 
In the SD phase near the Fermi level, there is also a decrease in DOS as the degeneracy of bands is lowered along the high symmetry k-paths.

\section{SHC with CDW in ScV$_6$Sn$_6$} \label{app:C}

\begin{figure}
\includegraphics[width=0.8\linewidth]{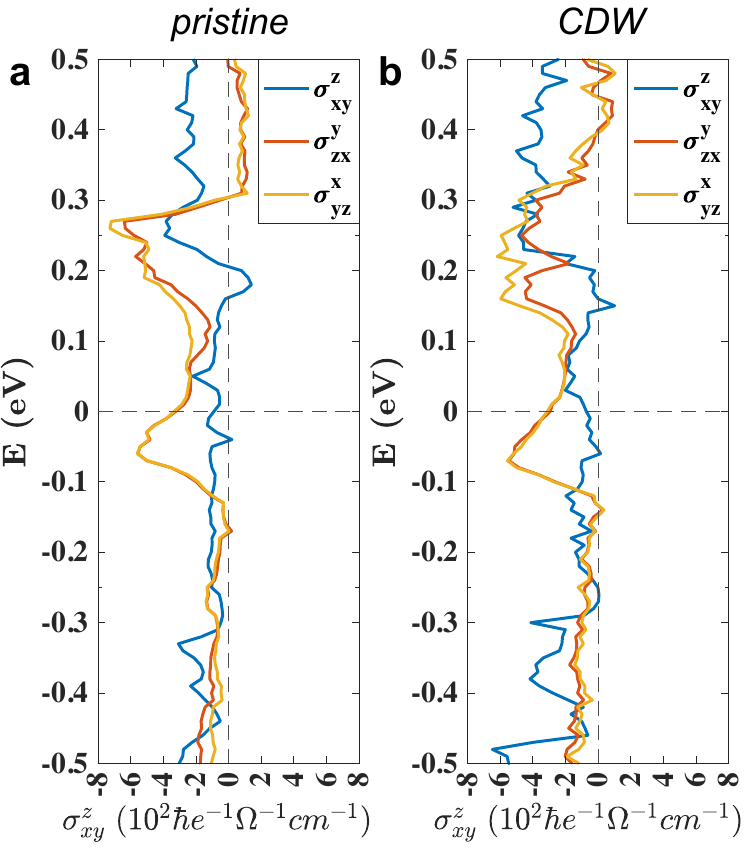}% Here is how to import EPS art
\caption{\label{fig:comp_SVS} Comparison of SHC components in ScV$_6$Sn$_6$ without (a) and with CDW (b).}
\end{figure}

Based on experimental observation~\cite{Arachchige2022charge}, ScV$_6$Sn$_6$ differs from CsV$_3$Sb$_5$ in terms of CDW distortion, with the former's unit cell expanding threefold in the $xy$-plane and the $z$ direction, resulting in a nine times larger unit cell volume. Besides the dissimilar ordering wave vectors, the two compounds also differ in the displacements of Vanadium atoms, with ScV$_6$Sn$_6$ displaying weaker displacements compared to the Star of David and Inverse Star of David in CsV$_3$Sb$_5$. Furthermore, in ScV$_6$Sn$_6$, Sc and Sn1 exhibit the strongest displacements, primarily in the $z$ direction.

We performed the SHC calculation for ScV$_6$Sn$_6$ in its CDW state, using the same method as described for CsV$_3$Sb$_5$ in the main text. The outcome, as depicted in Fig.~\ref{fig:comp_SVS}, shows negligible variation between the pristine phase and the CDW phase, signifying that the CDW in ScV$_6$Sn$_6$ does not substantially influence the spin response. This is not surprising, since the kagome sublattice is not prominently distorted in the CDW state.

%\bibliography{apssamp}
%apsrev4-2.bst 2019-01-14 (MD) hand-edited version of apsrev4-1.bst
%Control: key (0)
%Control: author (8) initials jnrlst
%Control: editor formatted (1) identically to author
%Control: production of article title (0) allowed
%Control: page (0) single
%Control: year (1) truncated
%Control: production of eprint (0) enabled
\providecommand{\noopsort}[1]{}\providecommand{\singleletter}[1]{#1}%

\end{document}